\documentstyle[prd,aps,epsfig]{revtex}
\def\be{\begin{equation}}
\def\ee{\end{equation}}
\begin{document}
\draft
\title{ Ultraviolet cut off, black hole-radiation equilibrium and
big bang.}
 
\author{Musongela Lubo \footnote{E-mail muso@ictp.trieste.it} \\
        The Abdus Salam International Centre for Theoretical Physics P.O.Box
586\\
34100 Trieste, Italy}
  
\maketitle
\begin{abstract}

In the presence of a minimal uncertainty in length, there exists a critical 
temperature above which the thermodynamics of a gas of radiation changes 
drastically.

We find that the  equilibrium temperature of a system composed 
 of a  Schwarzschild black hole surrounded  by radiation is unaffected by these
 modifications. This is  in agreement with works related to the robustness
 of the Hawking evaporation. The only change the deformation introduces concerns
 the critical volume at which the system ceases to be stable. 
 
 On the contrary, the evolution of the very early universe is sensitive to 
 the new behavior. We  readdress the shortcomings of the standard big
  bang model(flatness, entropy and horizon problems) in this context, assuming
  a minimal coupling to general relativity.
  Although they are not solved, some qualitative differences set in.
\end{abstract}
\section{Introduction}

The ultimate structure of space time has been at the core of many works. Some
begin with a fundamental construction like string theory and find that when 
particular fields are turned on, the effective theory can be described 
as built on a space time which has modified commutation relations 
\cite{LABELv1,LABELv2,LABELv3} or dispersion relations 
\cite{LABELv4}. The same occurs with  loop quantum gravity \cite{LABELv5}.
Another approach consists in using toy models with ad hoc modifications
in order to study, in a simplified way, the  influence of  possible
departures from usual symmetries at high scale. This last approach has been
adopted in the study of the  trans-Planckian problem of  black hole evaporation
\cite{Un2,Un1,Jac,co,brgr,LABEL32,LABELl2}. Our work fits in the second approach although
the commutators on which we work can be seen as coming from string theory 
\cite{LABELv2}.

\par  If one modifies the
commutators, one changes  the Heisenberg
uncertainty relations. The measure on the phase space is no more
the same; this results in new partition functions and
consequently different thermodynamical behaviors. From the quantum point of
view, the energy spectra of systems are modified by the change in
the commutation relations.

The influence of this "new" thermodynamics
in the early universe has been analyzed
in some models with  modified dispersion relations
\cite{LABEL21,ni2}. However, the equations of states used came only from
considerations on bosons. In the unmodified theory this is
justified by the fact that the difference between bosons and fermions reduces to
a factor $7/8$  in the energy densities, pressures, etc. As pointed out in
\cite{0305}, the difference between bosons and fermions in theories with
ultraviolet cut offs is much more pronounced. This naturally raises the question
of the way the picture is modified when one consider them altogether.
 In this work, we  provide analytical approximations for the equation of state,
the entropy of such a mixture and we quantify the flatness, entropy and 
 horizon problem in this framework. Many studies have been devoted to cosmological perturbations in transplanckian
physics \cite{mb,bm,iap,nie,np,tanaka,kempf,gree1,EGK01,mbk,shanka1,shanka2}; our treatment
tackles some of the reasons which led to the inflationary paradigm.
 
Concerning black holes, It  was  first  realized that, on
purely classical grounds, an entropy and a temperature could be associated to 
these objects 
\cite{Bek}. This was confirmed using Q.F.T on a curved background; the exact 
factor for the temperature was also found \cite{Haw}. It was then realized that 
in this derivation, photons were emitted with transplanckian frequencies, raising
doubts about their treatment as non interacting particles. It was pointed out 
afterward that taking somehow into account this effect trough  dispersion relations which depart
from the usual theory at transplanckian scale, the Hawking radiation was not
intrinsically changed while its derivation became more reliable 
 \cite{Un2,Un1,Jac,co,brgr}.   

When the dispersion relation of the photons is  the usual one, an equilibrium 
can be achieved for a system in which a neutral, non rotating and non charged 
black hole is in a fixed box filled with radiation \cite{dgp1,dgp2}. Moreover,
this temperature is the one obtained by Hawking. We analyze how this is 
affected when non trivial dispersion relations are considered, in the spirit 
of \cite{Un2,Un1,Jac,co,brgr}.

\par The article is organized as follows. In the second section we 
 briefly present a model exhibiting a minimal uncertainty in length and
derive its black-body radiation. We find sensible differences between fermions and bosons
at very high temperatures \cite{0305}. The third section treats  a system
consisting of a black hole in equilibrium with radiation in the new framework.
The last part investigates, quantitatively, how the problems of the
standard big bang(flatness, horizon, entropy) are affected.

\section{Black body radiation}
 
A very simple 
modification of the position-momentum commutation relation  leads to a 
theory possessing a minimal uncertainty in length \cite{LABEL1}. 
Some  high dimensional extensions  of this algebra  preserve  rotational
and translational invariance. The model we shall study is given by the 
following non vanishing commutators:
\be
\label{eq3}
[ \hat x_j, \hat p_k ] = i \hbar \left( f({\hat p}^2) \delta_{j k} + 
 g({\hat p}^2) \hat p_j \hat p_k \right) 
 \quad , \quad g(p^2) = \beta \quad , \quad
 f(p^2) = \frac{\beta p^2}{-1+\sqrt{1+ 2 \beta p^2}} \quad .
\ee
This theory has no position representation; the best way to recover information
on positions is trough the so called quasi-position representation
in which the momentum operators read:
\be
\label{eq4}
p_k = - i \hbar \sum_{r=0}^{\infty}  \left( \frac{\hbar^2 \beta}{2} 
\Delta \right)^r \frac{\partial}{\partial \xi_{k}} \quad , \quad {\rm where} 
\quad
\Delta = \sum_{l=1}^3  \frac{\partial^2}{\partial \xi_{l}^2} \quad .
\ee

Introducing the momentum scale $ \beta$ , it is straightforward
that with the Boltzmann constant $k$ , the light velocity $c$,
one can construct on purely
dimensional grounds the characteristic temperature 
\be
\label{eq5}
 \quad T_c =  \frac{c}{k
\sqrt{\beta}} \quad . \ee 

Let us now analyze how radiation gets affected by the new scale 
We will use the conventions of \cite{LABEL013,LABEL014}. Thanks to the 
deformation of the Klein-Gordon equation \cite{gordon}, the dispersion
relation in our case reads:
\be
\label{j1}
E = \frac{c \hbar k}{\left( 1+ \frac{1}{2} \hbar^2 \beta k^2 \right)}  
\quad .
\ee
The action of the momentum operators (Eq.(\ref{eq4})) on  plane waves
of wave vectors $\vec k$ is finite only if the condition
\be
\label{j2}
\hbar^2 k^2 \leq \frac{2}{\beta}
\ee
is satisfied; this is our cut off. The important quantities are
\be
\label{eq77}
q_{bo} = \sum_{\vec l} \log{\left( 1 - \exp{\left(- \frac{\epsilon_{\vec l}}{k T}
 \right)} \right)} = - \log{ Z_{bo}} \quad , \quad
 q_{fe} = - \sum_{\vec l} \log{\left( 1 + \exp{\left(- \frac{\epsilon_{\vec l}}{k T}
 \right)} \right)} = - \log{Z_{fe}} \quad ,
\ee
where $ Z$ is the grand partition function.
The entropy is given by
\be
\label{eq78}
S = - \frac{\partial \Phi}{\partial T} \quad , \quad {\rm with} \quad 
\Phi = k T \log Z \quad , \quad
\ee
while the energy and the particle number read
\be
\label{eq79}
 \quad
U = \sum_{\vec l}  \frac{\epsilon_{\vec l}}{ \exp{\left(
 \frac{\epsilon_{\vec l}}{k T} \right) \mp 1}} \quad , \quad
N = \sum_{\vec l}  \frac{1}{ \exp{\left(
 \frac{\epsilon_{\vec l}}{k T} \right) \mp 1}}  \quad .
\ee
For bosons, the quantity $q$ linked to the partition function by
Eq.(\ref{eq77}) can be written as
\be
\label{eq80}
q = 4 \pi V \left( \frac{k T}{h c}  \right)^3  \quad
\int_0^{\sqrt{2} \frac{T_c}{T} }  dx \,  x^2 
\log{ \left[ 1 - \exp{ \left( - 
\frac{x}{ 1 + \frac{1}{2} \frac{T^2}{T_c^2} x^2 } 
\right)} \right] } \quad ,
\ee
while the energy assumes the following form
\be
\label{eq81}
U = 4 \pi V  \frac{( k T)^4 }{(h c)^3}   \quad
\int_0^{\sqrt{2} \frac{T_c}{T} }  dx \, 
\frac{x^3}{ 1 + \frac{1}{2} \frac{T^2}{T_c^2} x^2} 
\left[  \exp{ \left(  \frac{x}{ 1 + \frac{1}{2} \frac{T^2}{T_c^2} x^2 } 
\right)} - 1  \right]^{-1} \quad .
\ee
The particle number  admits a similar  integral expression.

For temperatures  greater than or comparable to $T_c$, the interval of 
integration is small and a Taylor expansion  can be used to obtain an 
approximation. Working to fourth order, we are
led to the following expressions:
\begin{eqnarray}
\label{eq82}
p_{bo} &=& \sigma T_c \left[2 - \frac{2}{15} \sqrt{2}
 \frac{ T_c}{T}
 +  \sqrt{2}  \frac{ T}{T_c} \left( \frac{8}{3} \log{\frac{T}{T_c}} +
\frac{112}{45} - \frac{4}{3} \log{2} \right) \right]
\quad , \nonumber\\
 \rho_{bo} &=& \sigma T_c \left[ - 2 +
 \frac{8 }{3} \sqrt{2} \frac{ T}{T_c} +  \frac{4 }{15} \sqrt{2}
 \frac{ T_c}{T} \right] \quad , \quad 
 s_{bo} = \sigma \left[ \frac{2}{15} \sqrt{2} \left( \frac{ T_c}{T} \right)^2 +
   \frac{2}{45} \sqrt{2}   \left( 60 \log{\frac{T}{T_c}} +
116 - 30 \log{2} \right)  \right] \quad ,  \nonumber\\
N_{bo} &=& \frac{\sigma}{k} V  \left[ \frac{1}{3} \frac{T_c}{T} - 
\frac{4}{3} \sqrt{2} + 6 
\frac{T}{T_c}  \right]  \quad , \quad {\rm where} \quad \sigma = 
\pi \frac{k^4 T_c^3}{h^3 c^3} \quad .
\end{eqnarray}
The corresponding quantities for fermions(except the pressure)  are dominated by
constants:
\begin{eqnarray}
\label{eq83}
p_{fe} &=& \sigma T_c \left[ \frac{2}{5} \sqrt{2} 
 \frac{T_c}{T}   - 2 +  \frac{8}{3}  \log{2} \sqrt{2}
 \, \frac{T}{T_c}  \right] \quad , \quad
\rho_{fe} = \sigma T_c  \left(  2 -  \frac{4}{5} \sqrt{2} \,
  \frac{T_c}{T}  \right) \quad , \nonumber\\
s_{fe} &=& \sigma \left[ \frac{8}{3} \sqrt{2} \log{2} - \frac{2}{5} \sqrt{2}
\left( \frac{T_c}{T} \right)^2 \right] \quad , \quad
N_{fe} =   \frac{\sigma}{k} V \left( \frac{4}{3}  \sqrt{2}  -
 \frac{T_c}{T}  \right) \quad .
\end{eqnarray}

The behavior of the energy is depicted in Fig2.
\begin{figure}
\includegraphics{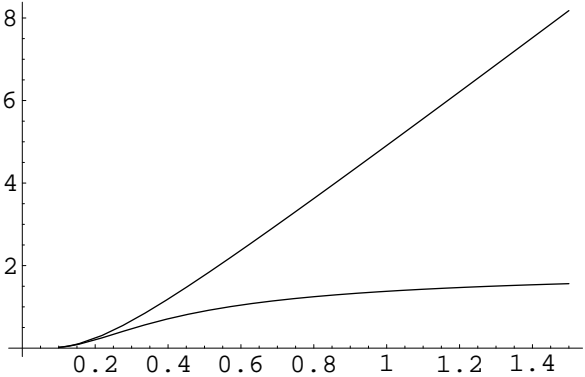}
\caption{The energy densities for fermions and bosons are plotted in 
terms of $\frac{T}{T_c}$. 
\newline
The unit for energies is 
$ 16 \pi  \frac{(k T_c)^4}{(h c)^3}  $. }
\end{figure} 
At temperatures below $T_c$, the energy density is polynomial ($ \sim T^4$).
Above $T_c$, it becomes linear as obtained in Eq.(\ref{eq82}) for bosons 
while it goes to a constant for fermions, as shown in Eq.(\ref{eq83}). The difference between bosons and fermions in the unmodified theory is encoded 
in the factor $7/8$, for the energy contributions for example. One sees this is 
drastically changed here.

\section{Black hole - radiation equilibrium}
 
 We have seen in the past section how thermodynamics is influenced by the 
 existence of a minimum uncertainty in length. We now wish to apply our results 
 to the only systems in which extremely high temperatures can be obtained:
  black holes and the very early universe. 
 
 Using purely classical considerations, it was argued by Beckenstein that the 
 area of a black hole can be interpreted as an entropy while its mass is 
 identified with the energy \cite{Bek}. The point of view considered in this
  paper is that the entropy of a black hole comes from a classical reasoning 
  and so is essentially the same as in the unmodified theory. This is linked
  to the fact that in most phenomenological approaches to trans-Planckian
  physics, one  suppose the particles evolve on a classical background
  but are subject to non trivial dispersion relations for example
  \cite{Un2,Un1,Jac,co,brgr,LABEL32,LABELl2}.  

Let us consider a Schwarzschild black hole  surrounded
by radiation. The entropy and the energy of such a system read: 
\be
\label{eq84}
S_{tot} = \frac{4 \pi}{l_{pl}^2} M^2 + S_{rad} \quad , \quad
E_{tot} = M c^2 + E_{rad} \quad ,
\ee   
where $l_{pl}$ is the Planck length and we use units in which $k=1$ \cite{fro}.
 If the system is isolated, the total energy is 
conserved. According to the second law of thermodynamics, 
equilibrium configurations correspond to maxima of the entropy. Therefore, if 
the volume of the system is fixed, the derivatives of $S_{tot} $ and $E_{tot}$
vanish at equilibrium. This can be used to obtain a relation between the mass of
the black hole and the equilibrium temperature: 
\be
\label{eq85}
M = \frac{l_{pl}^2 c^2}{8 \pi}  \left[  \frac{dS_{rad}}{dT}
\left( \frac{dE_{rad}}{dT}   \right)^{-1} \right]_{eq} \quad .
\ee

In the usual theory, one has
\begin{eqnarray} 
\label{eq86}
  \frac{dS_{rad}}{dT}
\left( \frac{dE_{rad}}{dT}   \right)^{-1}  &=& 
 \frac{d}{dT} \left( \frac{4 \pi^2}{45} \frac{ T^3}{c^3 h^3} \left(n_{bo}+ 
\frac{7}{8} n_{fe} \right) V \right)  \left[
 \frac{d}{dT} \left( \frac{ \pi^2}{15} \frac{ T^4}{c^3 h^3} \left(n_{bo}+ 
\frac{7}{8} n_{fe} \right)V \right) \right]^{-1} = \frac{1}{T} \nonumber\\  
& & \Longrightarrow
 T_{eq} = \frac{l_{pl}^2 c^2}{8 \pi} \frac{1}{M}   \quad ,
\end{eqnarray}
 so 
that the mass of a black hole is inversely proportional to its 
temperature; this is the Hawking temperature. Including fermions introduces 
$7/8$ factors but doesn't change the final result. 

This temperature is not 
affected by the presence of a minimal 
length uncertainty. To show this, let us first consider the regime in which a black hole is in 
equilibrium with  a radiation consisting uniquely of bosons in the extremely
high temperatures. Using the formulas displayed in Eq.(\ref{eq82}) and
 retaining only the 
dominant contributions one has
\be
\label{j3}
  \frac{dS_{rad}}{dT}
\left( \frac{dE_{rad}}{dT}   \right)^{-1}  =   
\frac{d}{dT} \left( \frac{8}{3} \sqrt{2} \log{\frac{T}{T_c}} \right) \left[
\frac{d}{dT} \left(\frac{8}{3} \sqrt{2} \frac{T}{T_c}  \right) \right]^{-1}= 
\frac{1}{T}  \quad .
\ee
Similarly, one has for a gas containing only fermions
\be
\label{j4}
  \frac{dS_{rad}}{dT}
\left( \frac{dE_{rad}}{dT}   \right)^{-1}  =   
\frac{d}{dT} \left( - \frac{2}{5} \sqrt{2} \frac{T_c^2}{T^2} \right) \left[
\frac{d}{dT} \left( - \frac{4}{5} \sqrt{2} \frac{T_c}{T}  \right) \right]^{-1}= 
\frac{1}{T}  \quad ,
\ee
so that for a mixture the temperature of equilibrium is unchanged.

This relation does not get corrections as one goes beyond the dominant contributions;
 it is  true  
at all temperatures. This can be seen explicitly for bosons by writing the integral related 
to the partition function in the following way:
\be
\label{eq93}
q =  4 \pi V \left( \frac{ T_c}{h c} \right)^3 \int_0^1 dy \, 2 \sqrt{2}
y^2  \,  \log{ \left(  1 - \exp{ (g(y))} \right) } \quad , \quad{\rm where} 
\quad
g(y) = - \sqrt{2} \frac{T_c}{T}   \frac{y}{1+y^2} \quad .
\ee 
Contrary to  Eq.(\ref{eq80}), the temperature does not appear in the upper bound
of the integral but in the integrand only. One can carry the derivative with respect
 to temperature trough the integral to obtain the entropy. The energy can in the same way be rewritten as
\be
\label{eq89}
E_{rad} = 16 \pi \,  V  \frac{ T_c^4}{(h c)^3}  \int_0^1
dy \frac{y^3}{1+y^2}  \frac{1}{\left[ \exp{(-g(y))} - 1 \right]} \quad .
\ee
Computing its derivative  one finds
\be
\label{energy}
\frac{dE_{rad}}{dT} = 4 \pi \sqrt{2} V \frac{ T_c^5}{c^3 h^3 T^2}
\int_0^1 \frac{y^4}{(1+y^2)^2}  \cosh^2{ \left( \frac{1}{\sqrt{2}} \frac{T_c}{T} 
\frac{y}{1+y^2} \right) } \quad .
\ee 
The derivative of the entropy 
\be
\label{deretr}
\frac{dS_{rad}}{dT} = -  \left( 2 \frac{\partial q}{\partial T} +
 T  \frac{\partial^2 q}{\partial T^2} \right)
\ee
is found to have  the same expression, with an extra factor $T$, so that
going back to 
Eq.(\ref{eq85})  the term under square brackets  is exactly $1/T$. A 
similar situation occurs for fermions. 

This is in agreement with the  idea that the Hawking black hole temperature
 measured by an observer at infinity is not affected by a modification
 of the dispersion relation at transplanckian energies \cite{Un2,Un1,Jac,co,brgr}.
 In fact, the result we just showed is generic and does not rely very much on
 the modified dispersion relation. The only think one needs is a fixed volume
 for the black hole- radiation system and a vanishing chemical potential for
 the particles composing the  radiation. In fact, using Eq.(\ref{eq77},\ref{eq78})
 without specifying any dispersion 
 relation, one obtains
 \be
 \label{eq90}
 \frac{dS_{rad}}{dT} = \frac{1}{k T^3}    
\sum_{\vec n} \epsilon_{\vec n}^2  \frac{ \exp{  
\left(- \frac{\epsilon_{\vec n}}{k T} \right) }}{
\left[ 1 - \exp{  \left(- \frac{\epsilon_{\vec n}}{k T} \right) } \right]^2} \quad .
 \ee
 Computing the derivative of the energy given in Eq.(\ref{eq79}) leads to a
 cancellation in Eq.(\ref{eq85}) which preserves exactly the last part 
 of Eq.(\ref{eq86}). This 
 reasoning is valid for any kind of "reasonable" dispersion relation i.e such
 that $q$ and its derivatives are finite and the derivation can be carried into 
 the infinite sum.
 
 There is a more direct way of obtaining the same result. The first law 
 of thermodynamics relates in the following way the changes in the energy $E_{rad}$, the
 entropy $S_{rad}$  
 and the number of particles $N$:
\be
\label{eq91}
dE_{rad} = T \, dS_{rad} - p \, dV + \mu \, dN \quad ,
\ee
$T$ being the temperature, $p$ the pressure and $\mu$ the chemical potential. 
The gas of radiation we consider is contained in a fixed volume($dV=0$) and has
 zero chemical potential($\mu=0$); this gives 
 \be
 \label{eq200}
 \frac{dS_{rad}}{dE_{rad}} = \frac{1}{T} \quad ,
 \ee
so that the term under  square brackets in Eq.(\ref{eq85}) takes the value $1/T$.

The equilibrium is stable if the second derivative of the entropy is positive.
This gives a particular volume such that the system tends
to a black hole  surrounded by radiation below it and to pure 
radiation above it. In our case, this critical
 volume is
 \be
 \label{vo}
V_o =  \frac{15 c^7 h^3 l_{pl}^2 M^2}
{\sqrt{2}    T_c^3 \left((5 c^4 l_{pl}^4 - 32 \pi^2 T_c^2 M^2) n_{bo} + 96 \pi^2
 T_c^2 M^2 n_{fe} \right)} ,
 \ee
where again we have put $k=1$.
Contrary to the equilibrium temperature which does not feel the modified 
dispersion relation, the volume which fixes the final fate of the system 
 depends on it. In particular, one sees that it goes like the second
power of the black hole mass, contrary to the usual case where it behaves like
its fifth power. Small black holes are the ones which display higher 
temperatures. The preceding formula is applicable only to such black holes.
The coefficients of $n_{fe}$ and $n_{bo}$ are thus positive and $V_o$ never 
blows up.
    
\section{The very early universe}
  
 The influence of transplanckian physics on the CMB predictions  
 has been the subject of many studies  
 \cite{mb,bm,iap,nie,np,tanaka,kempf,gree1,EGK01,mbk}. It was also pointed out 
   that the existence of a new physical scale changes the 
 equation of state for radiation and thus the evolution of the cosmic scale 
 factor \cite{LABEL21,ni2}. To our knowledge, all these studies  restricted 
 themselves  to the bosonic case. We showed in section $2$ that fermions, compared 
 to bosons have a drastically different behavior above $T_c$. 
 We will include them to the picture and quantify the horizon,
 entropy and flatness problem in this context.

We consider a Robertson-Walker space time
\be
\label{eq99}
ds^2 = - c^2 dt^2 + a(t)^2   \left( \frac{dr^2}{1 - K r^2} + r^2 
(d\theta^2 + \sin^2
\theta d\phi^2 )  \right) \quad ;
\ee
the constant $K$ can take the values  $1,0,-1$, corresponding
 respectively to the closed, flat,open  cases. 
 
To use general relativity one needs to restore covariance somehow. One of the
ways to tackle this issue is by introducing a unit vector field $u_\mu$ which is 
timelike and fixes the reference frame in which the modified dispersion relation 
hold. This field is made dynamical by the incorporation of a Maxwell type
Lagrangian for example. A multiplier is also used to impose the norm of this vector to be
equal to one. In a cosmological solution, this vector depends only on time. For
a theory exhibiting a modified dispersion relation of the form 
$E^2 = p^2 + a p^4$, a coupling to general relativity has been constructed in 
which the vector field  does not contribute to the energy-momentum tensor while
 the multiplier vanishes. The details of this procedure can be found  in 
 \cite{LABEL21}. Such a coupling will be said to be minimal. For simplicity, we
  shall assume such a coupling can be constructed
 in our model. The possibility of more subtle kinetic terms for the vector 
 field is also possible. If one chooses another coupling to gravity our analysis can be 
 seen as giving a heuristic
view  of the fact that  the bosonic and fermionic content  of the universe
has, in the presence of an ultraviolet cut off, an influence which is different 
from the one to which we are accustomed. The same thing holds if ultimately one
is not able to build the minimal coupling to gravity we  suggested above.

We shall then use the following equations for the evolution of the universe: 
\be
\label{eq94}
\dot \rho + 3 \frac{\dot a}{a} (\rho + p) = 0 \quad , \quad 
\left( \frac{\dot a}{a} \right)^2 = \frac{8 \pi G}{3 c^2} \rho + \frac{K}{a^2}
\quad ;
\ee
 the dot
means a derivative with respect to time. An important characteristic displayed by the thermodynamics of the system 
endowed with a minimal uncertainty in length  is that the 
equations of states 
 (see Eq.(\ref{eq82},\ref{eq83})) are not of the type $p= \gamma \rho$, with
 $\gamma$ a constant. This class of equations of states is common
 in cosmology  and leads to the  scale factor dependence $a \sim t^{\frac{2}{3} 
 \frac{1}{1+\gamma} }$. Therefore, in our case, the derivation of the relation between  the scale factor, 
time and temperature requires a slightly different treatment. Introducing a
 prime to denote the derivative with respect to temperature, the first part of
  Eq.(\ref{eq94}), with $K=0$, is transformed 
into the equation
\be
\label{eq95}
\frac{a^\prime}{a} = - \frac{\rho^\prime(T)}{p(T) + \rho(T)} \quad ,
\ee
which  admits a solution by quadratures:
\be
\label{eq96}
a(T) = a_{\star}  \exp{ \left[ -\frac{1}{3} \int_{T_\star}^T  
d\xi \, \frac{\rho^\prime(\xi)}{p(\xi) +
 \rho(\xi)}  \right] } \quad .
\ee 
As one knows the expressions for density and pressure in terms of the temperature, one
can infer the scale factor dependence on temperature. From the second
part of Eq.(\ref{eq94}), one reads similarly the link between time and
temperature:
\be
\label{eq97}
t(T) - t_\star = - \frac{1}{3} \sqrt{\frac{3 c^2}{8 \pi G}} \int_{T_\star}^T  
d\xi \, \frac{\rho^\prime(\xi)}{ \sqrt{\rho(\xi)} (p(\xi) +
 \rho(\xi))}  \quad .
\ee

In the usual case, one has $p = c^{te} T^4$ and $\rho = 3 p$. The last two 
equations then
give 
\begin{eqnarray}
\label{eq100}
a_2(T) &=& \tilde a_2 \frac{T_{pl}}{T} \quad  {\rm and} \quad 
 t_2(T) = \tilde t_2  \left( \frac{T_{pl}}{T} \right)^2 + \tilde t_3
\quad , \quad \tilde t_2 = \frac{3 \sqrt{5}}{4 \pi \sqrt{\pi}}  
\frac{1}{\sqrt{n_{bo}+ \frac{7}{8} n_{fe}}} t_{pl} \quad , \nonumber\\
& & {\rm where} \quad T_{pl} = \frac{1}{k}   \sqrt{\frac{c^5 \hbar}{G}}
\quad {\rm and} \quad t_{pl} =  \sqrt{\frac{G \hbar}{c^5}} 
\end{eqnarray}
are the Planck temperature and time.
From this one finds the usual relation $ t \sim 1/T^2 $.

The reason of the subscript for the scale factor is the following.
The  big bang in this model
has a radiation period composed of two stages: the first one corresponds to
extremely high temperatures and feels the 
presence of the minimal uncertainty in length. The evolution of the scale 
factor during that epoch will be denoted $a_1(t)$ and  given in
 Eqs.(\ref{eq98}). The second period is the one in which
the usual theory becomes valid; its evolution is given by Eqs.(\ref{eq100}) and its scale factor
is denoted $a_2(t)$.

At very high temperatures, bosons dominate. The evolution of the
scale factor  in terms of the temperature is then essentially given by the following parametric relations
\be
\label{ajout6}
t \sim  \left( \frac{T}{T_c} \right)^{-1/2}  \left(  \log{\frac{T}{T_c}} 
\right)^{-1}  \quad , \quad a 
 \sim \left(  \log{\frac{T}{T_c}} \right)^{-1/3} \quad .
\ee
Compared to the unmodified theory, we find that the scale factor's
evolution in terms of temperature is much slower. To put it differently,
at the same temperature, the usual big bang would predict a much bigger scale
factor, for temperatures well above $T_{c}$. The time spent to attain this
temperature is also more important in our model.
  
For future use, we will need the behavior at high temperatures but
at an epoch when fermions begin to play a role.
Eqs.(\ref{eq82},\ref{eq83},\ref{eq96},\ref{eq97}) lead to the following 
formulas:
\begin{eqnarray}
\label{eq98}
a_1(T) &=& \tilde a_1   \left[ \bar a \left(\frac{T}{T_c}\right)^{-2}
+ 2 \left( \bar d + \bar c \, \log{\frac{T}{T_c}}  \right) \right]^{-1/3} \quad , \nonumber\\
t_1(T) &=& \tilde t_1   \int_{T/T_c}^\infty  dx
\frac{( \bar c x^2 - \bar a )}{\sqrt{x} \sqrt{\bar c x^2+ \bar b x + \bar a}
\left[ \frac{\bar a}{2} + (\bar d + \bar c \log{x}) x^2 \right]} \quad , \quad
{\rm with} \quad \tilde t_1 = \frac{1}{\sqrt{3 \pi}} \left(\frac{T_{pl}}{T_c}\right)^2 
t_{pl} \quad .
\end{eqnarray}
The constants $\bar a, \cdots \bar d$,  embody the dependence of the system on the
bosonic-fermionic content at very high temperatures:
\begin{eqnarray}
\label{eq40}
\bar a &=& \frac{4}{15} \sqrt{2} (n^I_{bo} - 3 n^I_{fe} ) \quad , \quad
\bar b = 2  ( - n^I_{bo} +  n^I_{fe} )  \quad , \quad 
 \bar c =  \frac{8}{3} \sqrt{2} \, n^I_{bo} \quad , \nonumber\\
 \bar d &=& \frac{2}{45} \sqrt{2}  \left( 116\, n^I_{bo} - 30\,
   \left( n^I_{bo} - 2\, n^I_{fe} \right) \,
   \log{2}  \right) .
\end{eqnarray}
We have used a superscript ($n^I_{bo}$) to emphasize that the degrees of freedom appearing
here are the ones present at $T> T_c$.

The two periods must join smoothly at some point. For illustrative purposes, 
we shall make the approximation that this occurs at the critical temperature: we
 shall equalize the scale factors($a_1=a_2$) and the 
times($t_1=t_2$) at this
 value. The first equation leads to a relation between the constants fixing
 the scales of the geometries in the two regions:
\be
\label{eq41}
\frac{\tilde a_2}{\tilde a_1} = \left(  \bar a + 2  \bar d  \right) \frac{T_c}{T_{pl}} \quad ,
\ee
while the second can be used to extract the value of $\tilde t_3$.

\subsection{The flatness/entropy problem}

 The critical density $\rho_{cr}$  is a fictitious value which
 gives the same evolution of the scale factor but with $K=0$ in formula 
 (\ref{eq94}). Introducing $\Omega = \rho/\rho_{cr}$, one has 
 \be
 \label{eq101}
 \Omega - 1 = \frac{K}{ a^2 H^2 } \quad .
 \ee
 In the usual theory, one can use Eq.(\ref{eq100}) to show that
\be
\label{eq43}
\vert \Omega - 1   \vert_T = 
4 \left( \frac{\tilde t_2}{\tilde a_2} \right)^2 \left( \frac{T_{pl}}{T} 
\right)^2  \quad ,
\ee
from which one deduces 
 \be
 \label{eq102}
 \frac{\vert \Omega - 1 \vert_{T}}{\vert \Omega - 1 \vert_{T_{o}}} = 
\left( \frac{T_o}{T} \right)^2 = 10^{-64} \quad {\rm for} \quad T= T_{pl} 
\quad .
 \ee
In this formula $T_o$ is  the present day temperature of
the CMB radiation and $T_{pl}$ is the Planck temperature. As the  denominator 
of the left hand side of the preceding formula is known to be  close to
unity today, its numerator must have been incredibly close to one at the Planck scale:
this is the flatness problem; it can be solved by inflation.

Retaining the dominant contributions in Eqs.(\ref{eq98}), one obtains for the
first part of the radiation  period
\be
\label{42}
\vert \Omega - 1   \vert_T = 9  \, 2^{2/3}  (\bar c)^{-1/3}  
\left( \frac{\tilde t_1}{\tilde a_1} \right)^2
\left( \frac{T}{T_c} \right)^{-1}  \left( \log{ \frac{T}{T_c}} \right)^{2/3} 
 \quad ,
\ee
while Eq.(\ref{eq43}) is valid for the second period. As we don't have explicitly 
the scalar factor time dependence, we have obviously used the equation
\be
\label{eq45}
H = \frac{\dot a}{a} = \frac{1}{a} \, \frac{a'}{t'} \quad .
\ee

Using the matching condition displayed in Eq.(\ref{eq41}), one finds the
 ratio
\be
\label{eq44}
 \frac{\vert \Omega - 1 \vert_{T}}{\vert \Omega - 1 \vert_{T_{o}}} = Z
\left(  \frac{T_{o}}{T_c}\right)^{2}
\left(  \frac{T}{T_c} \right)^{-1}  
\left( \log{  \frac{T}{T_c} }  \right)^{2/3} \quad , \quad {\rm where} \quad
Z = \frac{4}{15} 2^{2/3} (\bar c)^{-1/3} \left( \bar a + 2 
\bar d  ) \right)^2 \left(n_{bo}^{II} + \frac{7}{8} 
n_{be}^{II} \right)
\quad .
\ee
We have used a  superscript ($n_{bo}^{II}$)  to emphasize that the particles appearing in 
this formula are the degrees of freedom present at $T_o$; their numbers are
of course smaller than the ones at the high temperature $T$ which enter into
the constants $\bar a, \cdots $ as displayed in Eq.(\ref{eq40}). 

From the last equation one draws two conclusions. First, as the temperature goes 
to infinity, the ratio vanishes so that the flatness problem is not
solved in this context. This was remarked 
in a different model \cite{LABEL21}, relying on numerical computations and in 
\cite{ni2} in the presence of a minimal uncertainty in length. The analytical approach followed in our work enables us to say two more things.
First, although the flatness problem is not solved, Eq.(\ref{eq44}) shows
that it is less severe in the presence of a minimal 
uncertainty in length. Second, the content of the theory in terms of bosons and 
fermions now plays a role, contrary to the usual theory in which Eq.(\ref{eq102})
applies. One also sees that assuming the critical and the Planck temperatures
to be different or equal doesn't matter; above the maximum of the two, our 
formulas show that the flatness  problem remains unsolved. 

At very high temperatures, one can rewrite the departure from flatness in terms 
of the entropy in the following way:
\be
\label{eq46}
\vert \Omega - 1 \vert_T \sim
 \frac{1}{s^{2/3} \exp{\left( \frac{3 s}{8 \sqrt{2} \sigma }\right)}} \quad ;
\ee
this differs from the undeformed theory for which one has  
\be
\label{eq47}
\vert \Omega - 1 \vert_T \sim
 \frac{1}{s^{2/3}}  \quad .
\ee
The same conclusion than in the unmodified theory holds: to be almost flat 
today, the universe had to have a huge entropy density at the initial times.

In our model, one can verify that the total entropy is conserved, order by 
order. For example, the dominant contribution to the entropy contained in a
comoving volume is
\be
\label{ajout10}
a^3 s \sim  c^{st} \left[ \left( \log{ \frac{T}{T_c}} \right)^{-1/3} \right]^3 
\left( \log{ \frac{T}{T_c}} \right)
\ee
which is independent of the temperature. 

\subsection{The horizon problem}

The distance a photon has traveled since the big bang is given by
\be
\label{eq49}
R_H = c \, a(t_o) \int_0^{t_o}  \frac{dt}{a(t)} = - c \, a(t_o) \left[ 
\int_{T_o}^{ T_c}  t_2'(T) \frac{d T}{a_2(T)} + 
\int_{ T_c}^{\infty}  t_1'(T) \frac{d T}{a_1(T)}   
\right] \quad .
\ee 
The dominant contributions can be written as
\be
\label{eq50}
R_H(T_o) = R_H^\star(T_o) \left( 1 + \gamma \frac{T_o}{T_c} \right)
  \quad ,
\ee
where $ R_H^\star(T_o)$ is the horizon in the unmodified theory while $\gamma$
contains the information about the fermionic/bosonic content:
\be
\label{eq51}
\gamma = 2^{-2/3} \frac{4 \pi}{3 \sqrt{15}} \sqrt{n_{bo}^{II} + \frac{7}{8}
n_{fe}^{II} } \left(  \bar a +  2 \bar d    \right)
\int_{1}^\infty  dx
\frac{( \bar c x^2 - \bar a )}{ x^{7/8} \sqrt{\bar c x^2+ \bar b x + \bar a}
\left[ \frac{\bar a}{2} + (\bar d + \bar c \log{x}) x^2 \right]^{2/3}} \quad .
\ee 

The smallness of  today's temperature $T_o$ compared to the critical 
temperature $T_c$ is such that the correction to the horizon
will not be significant. Note, however that once again the spin content of the
universe enters into play in a non trivial way.

The various quantities which appear in our formulas have the correct behaviours.
 Considering for example Eq.(\ref{eq51}), the term under square root
in the integral must obviously be positive. One can rewrite it as a linear
combination of $n^{I}_{fe}$ and $n^{I}_{bo}$; the first coefficient vanishes at
$x= 0.56$ while the second has no zero. This means that the matching 
temperature can not be taken to be lower than $0.56 T_c$. A precise numerical 
evaluation of this quantity is possible but one  has then to work in a
specific model, with $n^{I}_{fe},n^{I}_{bo}$ known. One can nevertheless
expect that the 
departure from usual physics will take place around $T_c$.

\section{Conclusions}

We have studied the thermodynamics induced by a non local theory which 
exhibits a minimal uncertainty in length. We have obtained that a new behavior
sets in at very high temperatures. The difference between fermions and bosons is
more important than in the usual case. When studying the equilibrium of a gas of
radiation surrounding a black hole, we have suggested a generic argument which
 assures the universality of the temperature of equilibrium, for reasonable
deformations. Our work is in agreement with previous studies which, using the 
machinery of quantum field theory  in  curved backgrounds,
showed that the Hawking radiation as perceived by a remote observer is not
affected when the dispersion relation gets modified at Planckian energies.
The only difference we found so far is in the  volume which is at the 
frontier separating the cases in which the final stage contains only radiation
from the ones in which the two are present.
On the contrary, we have shown that the flatness problem is less sever in this
context, contrary to the horizon problem which remains roughly speaking
untouched. The scale factor and time growths in function of the temperature
are slower than in the usual big bang. We also saw that the entropy per 
comoving volume was conserved.

One of the important ingredients in our cosmological analysis is the way the 
theory is coupled to gravity. We assumed, inspired by 
\cite{LABEL21}, that there is a "minimal" coupling i.e one for which, in the   
cosmological solution, the unit time-like  vector field does not contribute to the energy density. One of the challenges is now
to construct explicitly such a model. If one considers other possibilities,
quantitative differences are likely to appear. We nevertheless suspect that,
qualitatively, the fact that bosons and fermions behave differently at very
high temperatures is captured, at least partially, by the treatment we have presented. One of the
key issues in a more complete treatment will be the adiabatic expansion of the
universe.

\par The commutation relations studied
here can be interpreted as phenomenological consequences of string
or M theory \cite{LABELv2}. Our work, after others,
 suggests that string cosmology may not be uniquely characterized by the
evolution of the fields which appear at the lowest order(like the
dilaton) but also by some non trivial statistical effects. Finally, the
theory exhibiting a minimal length uncertainty may forbid the
singularity present in the standard big bang scenario. A similar
reasoning was advocated to argue that the Hawking evaporation of a
black hole may halt without using the complementarity hypothesis
\cite{LABEL34}.

\vskip 1cm

\underline{Acknowledgments}
 
I thank G.Senjanovic and  H.G.Casini for useful remarks. 

\end{document}